\documentstyle[iopconf2,psfig,epsfig,12pt]{article}

\def\bnue{\hbox{$\bar\nu_e$ }}  
\def\bnum{\hbox{$\bar\nu_\mu$ }}  
\def\bnut{\hbox{$\bar\nu_\tau$ }} 

\newcommand {\ignore}[1]{}

\newcommand{\bc}{\begin{center}}
\newcommand{\ec}{\end{center}}

\def\ifmath#1{\relax\ifmmode #1\else $#1$\fi}
\def\half{\ifmath{{\textstyle{1 \over 2}}}}

\def\3quarter{{\textstyle{3 \over 4}}}

\def\ra{\rightarrow}

\def\lf{\leaders\hbox to 1em{\hss.\hss}\hfill}
\def\e6{$E(6)$}
\def\10{$SO(10)$}
\def\21{$SU(2) \otimes U(1) $}
\def\lr{$SU(2)_L \otimes SU(2)_R \otimes U(1)$}
\def\422{$SU(4) \otimes SU(2) \otimes SU(2)$}
\def\321{$SU(3) \otimes SU(2) \otimes U(1)$}
\def\ne{\hbox{$\nu_e$ }}
\def\nm{\hbox{$\nu_\mu$ }}
\def\nt{\hbox{$\nu_\tau$ }}
\def\ns{\hbox{$\nu_{s}$ }}

\def\mnt{\hbox{$m_{\nu_\tau}$ }}

\def\eg{\hbox{\it e.g., }}        

\def\neus{\hbox{neutrinos }}
\def\gau{\hbox{gauge }}
\def\neu{\hbox{neutrino }}

\def\eq#1{{eq. (\ref{#1})}}

\def\fig#1{{Fig. (\ref{#1})}}

\def\VEV#1{\left\langle #1\right\rangle}

\def\lsim{\raise0.3ex\hbox{$\;<$\kern-0.75em\raise-1.1ex\hbox{$\sim\;$}}}
\def\gsim{\raise0.3ex\hbox{$\;>$\kern-0.75em\raise-1.1ex\hbox{$\sim\;$}}}
\def\half{{1\over 2}}
\def\beq{\begin{equation}}
\def\eeq{\end{equation}}
\def\bef{\begin{figure}}
\def\eef{\end{figure}}
\def\bet{\begin{table}}
\def\eet{\end{table}}
\def\bea{\begin{eqnarray}}
\def\ba{\begin{array}}
\def\ea{\end{array}}
\def\bi{\begin{itemize}}
\def\ei{\end{itemize}}
\def\ben{\begin{enumerate}}
\def\een{\end{enumerate}}
\def\ra{\rightarrow}

\def\eea{\end{eqnarray}}
\def\apj#1#2#3{          {\it Astrophys. J. }{\bf #1} (19#2) #3}

\def\jel#1#2#3{         {\it Journal Europhys. Lett. }{\bf #1} (19#2) #3}
\def\ib#1#2#3{           {\it ibid. }{\bf #1} (19#2) #3}
\def\nat#1#2#3{          {\it Nature }{\bf #1} (19#2) #3}
\def\nps#1#2#3{        {\it Nucl. Phys. B (Proc. Suppl.) }{\bf #1} (19#2) #3} 
\def\np#1#2#3{           {\it Nucl. Phys. }{\bf #1} (19#2) #3}
\def\pl#1#2#3{           {\it Phys. Lett. }{\bf #1} (19#2) #3}
\def\pr#1#2#3{           {\it Phys. Rev. }{\bf #1} (19#2) #3}
\def\prep#1#2#3{         {\it Phys. Rep. }{\bf #1} (19#2) #3}
\def\prl#1#2#3{          {\it Phys. Rev. Lett. }{\bf #1} (19#2) #3}
\def\pw#1#2#3{          {\it Particle World }{\bf #1} (19#2) #3}

\def\rmp#1#2#3{          {\it Rev. Mod. Phys. }{\bf #1} (19#2) #3}
\def\zp#1#2#3{           {\it Zeit. fur Physik }{\bf #1} (19#2) #3}

\def\n.c.#1#2#3{         {\it Nuovo Cim. }{\bf #1} (19#2) #3}
\def\r.n.c.#1#2#3{       {\it Riv. del Nuovo Cim. }{\bf #1} (19#2) #3}
\def\sjnp#1#2#3{         {\it Sov. J. Nucl. Phys. }{\bf #1} (19#2) #3}

\def\jetp#1#2#3{         {\it JETP }{\bf #1} (19#2) #3}
\def\mpl#1#2#3{          {\it Mod. Phys. Lett. }{\bf #1} (19#2) #3}

\def\ppnp#1#2#3{           {\it Prog. Part. Nucl. Phys. }{\bf #1} (19#2) #3}

\def\tp{these proceedings}
\def\pc{private communication}
\def\opc{\hbox{{\sl op. cit.} }}
\def\ip{in preparation}

\begin{document}
\def\bnue{\hbox{$\bar\nu_e$ }}  
\def\bnum{\hbox{$\bar\nu_\mu$ }}  
\def\bnut{\hbox{$\bar\nu_\tau$ }} 
\large{
\title{ Physics Beyond the Desert 
\footnote{Invited talk at the Workshop on Physics beyond the Standard Model
Accelerator- and Non-Accelerator approaches ("Beyond the Desert")
Castle Ringberg, Tegernsee, Germany, 8-14 June 1997.} }
\author{Jos\'e W. F. Valle\ddag}
\affil{\ddag Instituto de F\'{\i}sica Corpuscular 
- C.S.I.C.\\Departament de F\'{\i}sica Te\`orica, Universitat de
Val\`encia\\46100 Burjassot, Val\`encia, Spain\\
http://neutrinos.uv.es}
}
\beginabstract 

I review the observational status of neutrino physics, including the
present hints for \neu mass and the ways to reconcile the solar and
atmospheric neutrino data with the existence of a hot dark matter
component, and the possible hints from LSND.  I also briefly discuss
the electroweak symmetry breaking sector of the Standard Model (SM),
focussing on supersymmetric models with broken R--parity and
spontaneously broken lepton number. I discuss some of the signals
expected at future accelerators such as LEP II and LHC. They serve to
illustrate how neutrino mass effects may be testable not only at
underground and nuclear physics installations but also at high energy
collider experiments.

\endabstract

\thispagestyle{empty}

\newpage

\setcounter{page}{1}

\large{
\title{ Physics Beyond the Desert}
\author{Jos\'e W. F. Valle\ddag}
\affil{\ddag Instituto de F\'{\i}sica Corpuscular 
- C.S.I.C.\\Departament de F\'{\i}sica Te\`orica, Universitat de
Val\`encia\\46100 Burjassot, Val\`encia, Spain\\
http://neutrinos.uv.es}
}
\beginabstract 

I review the observational status of neutrino physics, including the
present hints for \neu mass and the ways to reconcile the solar and
atmospheric neutrino data with the existence of a hot dark matter
component, and the possible hints from LSND.  I also briefly discuss
the electroweak symmetry breaking sector of the Standard Model (SM),
focussing on supersymmetric models with broken R--parity and
spontaneously broken lepton number. I discuss some of the signals
expected at future accelerators such as LEP II and LHC. They serve to
illustrate how neutrino mass effects may be testable not only at
underground and nuclear physics installations but also at high energy
collider experiments.

\endabstract

\section{Introduction}
\vskip .1cm

Neutrinos are the only apparently massless electrically neutral
fermions in the SM and the only ones without \21 singlet partners.  It
is rather mysterious why \neus seem to be so special when compared
with the other fundamental fermions. Indeed one of the most unpleasant
features of the SM is that the masslessness of neutrinos is not
dictated by an underlying {\sl principle}, such as that of gauge
invariance in the case of the photon: the SM simply postulates that
neutrinos are massless and, as a result, many of their properties are
trivial.  If massive, neutrinos would present another puzzle, of why
are their masses so much smaller than those of the charged
fermions. The fact that neutrinos are the only electrically neutral
elementary fermions may hold the key to the answer, namely neutrinos
could be Majorana fermions, the most fundamental ones. In this case
the suppression of their mass could be associated to lepton number
conservation, as actually happens in many extensions of the SM.

Although attractive, the seesaw mechanism \cite{GRS} is by no means
the only way to generate \neu masses. There are many other attractive
possibilities, some of which do not require any large mass scale. The
extra particles required to generate the \neu masses have masses
accessible to present experiments \cite{zee.Babu88}.

It is also quite plausible that B-L or lepton number, instead of being
part of the gauge symmetry \cite{GRS,LR} may be a spontaneously broken
global symmetry. The scale at which such a symmetry gets broken does
not need to be high, as in the original proposal \cite{CMP}, but can be
rather low, close to the weak scale \cite{JoshipuraValle92}.  Such a
low scale for lepton number breaking could have important implications
not only in astrophysics and cosmology but also in particle physics.

This large diversity of possible schemes and the lack of a theory for
the Yukawa couplings imply that present theory is not capable of
predicting the scale of \neu masses any better than it can fix the
masses of the other fermions, like that of the muon. As a result one
should at this point turn to experiment.

From the observational point of view non-zero neutrino masses now seem
required in order to account for the data on solar and atmospheric
neutrinos, as well as cosmological data on the amplitude of primordial
density fluctuations which suggest the need for hot dark matter in the
universe.  I briefly over-view the present observational limits and
hints in favour of massive neutrinos, and make a few general remarks
about the theoretical models.

Turning to the electroweak breaking sector, I mention some of the
physics motivations and potential of various extensions of the
SM, with emphasis on supersymmetry (SUSY) with broken R--parity
and majoron extensions of the SM. Some of the related
phenomena are deeply related to the neutrino sector of the theory. I
discuss the some aspects of the physics of invisibly decaying Higgs
bosons and its impact on Higgs boson searches at accelerators, such as
LEP II.  Second, I discuss a few aspects of models beyond the Minimal
Supersymmetric Standard Model (MSSM) phenomenology, in which R parity
is violated, as well as some of the associated laboratory signatures.

Many of the new signatures may be accessible to experiments performed
at accelerators or at underground installations, thus illustrating the
complementarity between these two approaches in the search for signals
beyond the SM.

\section{Models of Neutrino Mass}
\vskip .1cm

One of the most attractive approaches to generate neutrino masses is
from unification. Indeed, in trying to understand the origin of parity
violation in the weak interaction by ascribing it to a spontaneous
breaking phenomenon, in the same way as the W and Z acquire their
masses in the SM, one arrives at the so-called left-right symmetric
extensions such as \lr \cite{LR}, \422 \cite{PS} or \10 \cite{GRS}, in
some of which the masses of the light neutrinos are obtained by
diagonalizing the following mass matrix in the basis $\nu,\nu^c$
\begin{equation}
\left[\matrix{
 0 & D \cr
 D^T & M_R }\right] 
\label{SS} 
\end{equation} 
where $D$ is the standard \21 breaking Dirac mass term and $M_R =
M_R^T$ is the isosinglet Majorana mass. In the seesaw approximation,
one finds
\beq M_{eff} = - D M_R^{-1} D^T \:.  
\label{SEESAW} 
\eeq 
In general, however, this matrix also contains a $\nu\nu$ term
\cite{2227} whose size is expected to be also suppressed by the
left-right breaking scale. As a result one is able to explain
naturally the relative smallness of \neu masses.  Even though it is
natural to expect $M_R$ to be large, its magnitude heavily depends on
the model. Moreover the $M_R$ may have different possible structures
in flavour space (so-called textures) \cite{Smirnov}.  As a result one
can not make any real prediction for the corresponding light neutrino
masses that are generated through the seesaw mechanism. In fact this
freedom has been exploited in model building in order to account for
an almost degenerate \neu mass spectrum \cite{DEG}.

Although very attractive, unification is not the only way to generate
neutrino masses. There are many other attractive schemes which do not
require any large mass scale. For example, it is possible to start
from an extension of the lepton sector of the \21 theory by adding a
set of $two$ 2-component isosinglet neutral fermions, denoted
${\nu^c}_i$ and $S_i$, to each generation $i$. In this case there is an
exact L symmetry that keeps neutrinos strictly massless, as in the
SM. The conservation of total lepton number leads to the following
form for the neutral mass matrix (in the basis $\nu, \nu^c, S$)
\begin{equation}
\left[\matrix{
  0 & D & 0 \cr
  D^T & 0 & M \cr
  0 & M^T & 0 }\right] 
\label{MAT} 
\end{equation} 
This form has also been suggested in various theoretical models
\cite{WYLER}, including superstring inspired models.  In the latter
case the zeros of \eq{MAT} naturally arise due to the absence of Higgs
fields to provide the usual Majorana mass terms, needed in the seesaw
scheme \cite{SST}. Clearly, one can easily introduce non-zero masses
in this model through a $\mu S S$ term that could be proportional to
the vacuum expectation value (VEV) of a singlet field $\sigma$
\cite{CON}. In contrast to the seesaw scheme, the \neu masses are
directly proportional to $\VEV{\sigma}$. This model provides a
conceptually simple and phenomenologically rich extension of the SM,
which opens the possibility of many new phenomena. These have to do
with neutrino mixing, universality, flavour and CP violation in the
lepton sector \cite{BER,CP}, as well as direct effects associated with
Neutral Heavy Lepton (NHL) production at high energy colliders
\cite{CERN}.  A remarkable feature of this model is the possibility of
non-trivial neutrino mixing despite the fact that neutrinos are
strictly massless. This tree-level effect leads to a new type of
resonant \neu conversion mechanism that could play an important role
in supernovae \cite{massless0,massless}.  Moreover, there are
loop-induced lepton flavour and CP non-conservation effects whose
rates are precisely calculable \cite{BER,CP,3E}. I repeat that this is
remarkable due to the fact that physical light neutrinos are massless,
as in the SM. This feature is the same as what happens in the
supersymmetric mechanism of flavour violation \cite{Hall}. Indeed, in
the simplest case of SU(5) supergravity unification, there are flavour
violating processes, like $\mu \ra e \gamma$, despite the fact that in
SU(5) neutrinos are protected by B-L and remain massless. The
SUSY mechanism and that of \eq{MAT} differ in that the lepton 
flavour violating processes are induced in one case by NHL
loops, while in SUSY they are induced by scalar boson loops.
In both cases the particles in the loops have masses at the weak
scale, leading to branching ratios \cite{BER,CP,3E}
\cite{SUSYLFV,SUSYLFV2} that are sizeable enough to be of experimental
interest \cite{ETAU,TTTAU,opallfv}.

\begin{figure}[t]
\centerline{\protect\hbox{\psfig{file=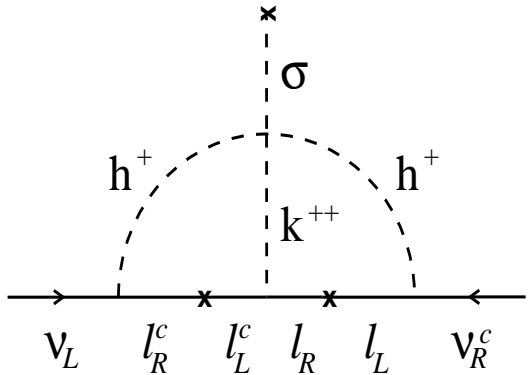,height=5cm,width=7cm}}}
\caption{Two-loop-induced Neutrino Mass. }
\label{2loop}
\end{figure}

There is also a large variety of {\sl radiative} schemes to generate
\neu masses. The prototype models of this type are the Zee model and
the model suggested by Babu \cite{zee.Babu88}. In these models lepton
number is explicitly broken, but it is easy to realize them with
spontaneous breaking of lepton number. For example in the version
suggested in ref. \cite{ewbaryo} the neutrino mass arises from the
diagram shown in \fig{2loop}.

\begin{figure}[t]
\centerline{\protect\hbox{\psfig{file=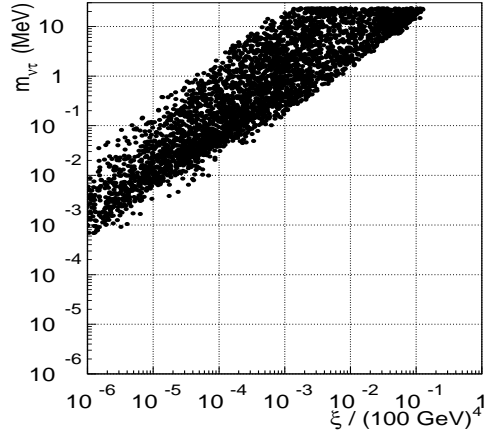,height=6cm,width=7cm}}}
\caption{Tau neutrino mass versus $\epsilon_3$ }
\label{mnt_xi_ev}
\end{figure}

\begin{figure}[t]
\centerline{\protect\hbox{\psfig{file=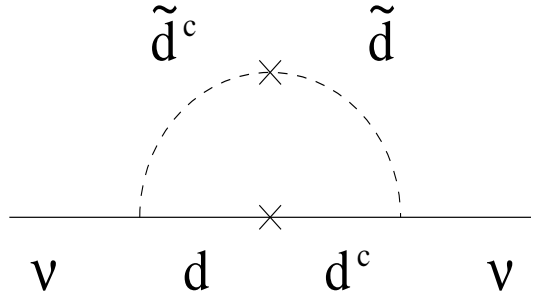,height=4.5cm,width=7cm}}}
\vglue -0.6cm
\caption{Scalar loop contributions to neutrino masses  }
\label{mnrad}
\end{figure}

The seesaw and the radiative mechanisms of neutrino mass generation
may be combined. Supersymmetry with broken R-parity also provides a
very elegant mechanism for the origin of neutrino mass, as well as
mixings \cite{epsrad}. Here I focus on the simplest unified
supergravity version of the model with bilinear breaking of R--parity,
characterized universal boundary conditions for the soft breaking
parameters \cite{epsrad,RPothers}. In this model the tau neutrino
$\nu_{\tau}$ acquires a mass, due to the mixing between \neus and
neutralinos given in the matrix
\begin{equation}
\left[\matrix{
M_1 & 0  & -\half g'v_1 & \half g'v_2 & -\half g'v_3 \cr
0   & M_2 & \half g v_1 & -\half g v_2 & \half g v_3 \cr
-\half g'v_1 & \half g v_1 & 0 & -\mu & 0 \cr
\half g'v_2 & -\half g v_2 & -\mu & 0 & \epsilon_3 \cr
-\half g'v_3 & \half g v_3 & 0 & \epsilon_3 & 0 
}\right]
\label{eq:NeutMassMat}
\end{equation}
This model contains only one extra free parameter in addition to those
of the minimal supergravity model, as the $\epsilon_3$ and the $v_3$
are related by a minimization condition. Contrary to a popular
misconception, the bilinear violation of R--parity implied by the
parameter $\epsilon_3$ is physical and can not be rotated away
\cite{DJV}. In fact, what happens in this model with universal
conditions for the soft breaking parameters is that the value of
$\epsilon_3$ is induced radiatively, due to the effect of the non-zero
bottom quark Yukawa coupling $h_b$ in the running of the
renormalization group equations from the unification scale down to the
weak scale \cite{DJV}.  This makes $\epsilon_3$ and \mnt {\sl
calculable}.  Thus \eq{eq:NeutMassMat} is analogous to a see-saw type
matrix \eq{SS}, in which the $M_R$ lies at the weak scale
(neutralinos), while the r\^ole of the Dirac entry $D$ is played by
the $\epsilon_3$, which is, in a sense, a radiatively induced
quantity. From this point of view, the mechanism is a {\sl hybrid}
see-saw like scheme, with naturally suppressed Majorana $\nu_{\tau}$
mass induced by the mixing between weak eigenstate neutrinos and {\sl
Higgsinos} or {\sl gauginos}. The \nt mass induced this way depends
quadratically on an effective parameter $\xi$ defined as $\xi
\equiv (\epsilon_3 v_1 + \mu v_3)^2$ characterizing the
violation of either through $v_3$ or $\epsilon_3$.  In \fig{mnt_xi_ev}
we display the allowed values of $m_{\nu_{\tau}}$ which clearly can be
quite low, due to the possible cancellation between the two terms in
$\xi$. In unified supergravity models with universal soft masses
this cancellation happens automatically and is, as mentioned,
calculable in terms of $h_b$. 
Notice that \ne and \nm remain massless in this approximation. They
get masses either from scalar loop contributions in \fig{mnrad}
\cite{RPnuloops,DJV}
or by mixing with singlets in models with spontaneous breaking of
R-parity \cite{Romao92}.  It is important to notice that even when 
\mnt is small, many of the corresponding R-parity violating effects 
can be sizeable. An obvious example is the fact that the lightest
neutralino decay will typically decay inside the detector, unlike 
the case of the MSSM.

Other than the seesaw scheme, none of the above models requires a
large mass scale. In all of them one can implement the spontaneous
violation of the global lepton number symmetry leading to \neu masses
that scale {\sl directly} proportional to the lepton-number scale or
some positive power of it, in contrast to the original majoron model
\cite{CMP}. Such low-scale models are very attractive and lead to a
richer phenomenology, as the extra particles required have masses at
scales that could be accessible to present experiments. One remarkable
example is the possibility invisibly decaying Higgs bosons
\cite{JoshipuraValle92}.

The above discussion should suffice to illustrate the enormous freedom
and wealth of phenomenological possibilities in the neutrino
sector. These reach well beyond the realm of conventional neutrino
experiments, including also signatures that can be probed, though
indirectly, at high energy accelerators.  An optimist would regard as
very exciting the fact that the neutrino sector may hold so many
experimental possibilities, while a pessimist would be discouraged by
the fact that one does not know the relevant scale responsible for
neutrino mass, nor the underlying mechanism. Last but not least, one
lacks a theory for the Yukawa couplings. As a consequence \neu masses
are not predicted and it is up to observation to search for any
possible clue.  Given the theoretical uncertainties in predicting \neu
masses from first principles, one must turn to observation. Here the
information comes from laboratory, astrophysics and cosmology.

\subsection{Laboratory Limits }
\vskip .1cm

The best limits on the neutrino masses can be summarized as
\cite{PDG96}:
\beq
\label{1}
m_{\nu_e} 	\lsim 5 \: \mbox{eV}, \:\:\:\:\:
m_{\nu_\mu}	\lsim 170 \: \mbox{keV}, \:\:\:\:\:
m_{\nu_\tau}	\lsim 18 \: \mbox{MeV}
\eeq
These are the most model-independent of the laboratory limits on \neu
mass, as they follow purely from kinematics.  The limit on the \ne
mass comes from beta decay, that on the \nm mass comes from PSI (90 \%
C.L.) \cite{psi}, with further improvement limited by the uncertainty
in the $\pi^-$ mass.  On the other hand, the best \nt mass limit now
comes from high energy LEP experiments \cite{eps95} and may be
substantially improved at a future tau-charm factory \cite{jj}. In
connection with tritium beta decay limit \cite{Lobashev} even though
the negative $m^2$ value has now been clarified, there are still
un-understood features in the spectrum, probably of instrumental
origin. Further results from the Mainz experiment are awaited.

Additional limits on neutrino masses follow from the non-observation
of neutrino oscillations.  The most stringent bounds come from reactor
oscillation experiments \cite{reactor} (\bnue - $\nu_x$ oscillations).
Here we highlight the recent results of the first long-baseline
reactor neutrino oscillation experiment Chooz \cite{Chooz}. There are
also stringent bounds from meson factory oscillation experiments
(KARMEN \cite{karmen}, LSND \cite{lsnd}) and from high-energy
accelerator experiments E531 and E776 \cite{E531.E776} (\nm - \nt).  A
search for \nm to \ne oscillations has now been reported by the LSND
collaboration using \nm from $\pi^+$ decay in flight
\cite{lsndflight}. An excess in the number of beam-related events from
the $C(\nu_e,e^-)X$ inclusive reaction is observed. The excess cannot
be explained by normal \ne contamination in the beam at a confidence
level greater than 99\%.  If interpreted as an oscillation signal, the
observed oscillation probability of $(2.6 \pm 1.0 \pm 0.5) \times
10^{-3}$ is consistent with the previously reported \bnum to \bnue
oscillation evidence from LSND. Another recent result comes from NOMAD
and rules out part of the LSND region. The future lies in searches for
oscillations using accelerator beams directed to far-out underground
detectors, with very good prospects for the long-baseline experiments
proposed at KEK, CERN and Fermilab.

If neutrinos are of Majorana type a new form of nuclear double beta
decay would take place in which no neutrinos are emitted in the final
state, i.e. the process by which an $(A,Z-2)$ nucleus decays to $(A,Z)
+ 2 \ e^-$. In such process one would have a virtual exchange of
Majorana neutrinos. Unlike ordinary double beta decay, the
neutrino-less process violates lepton number and its existence would
indicate the Majorana nature of neutrinos.  Because of the phase space
advantage, this process is a very sensitive tool to probe into the
nature of neutrinos.

\begin{figure}[t]
\centerline{\protect\hbox{\psfig{file=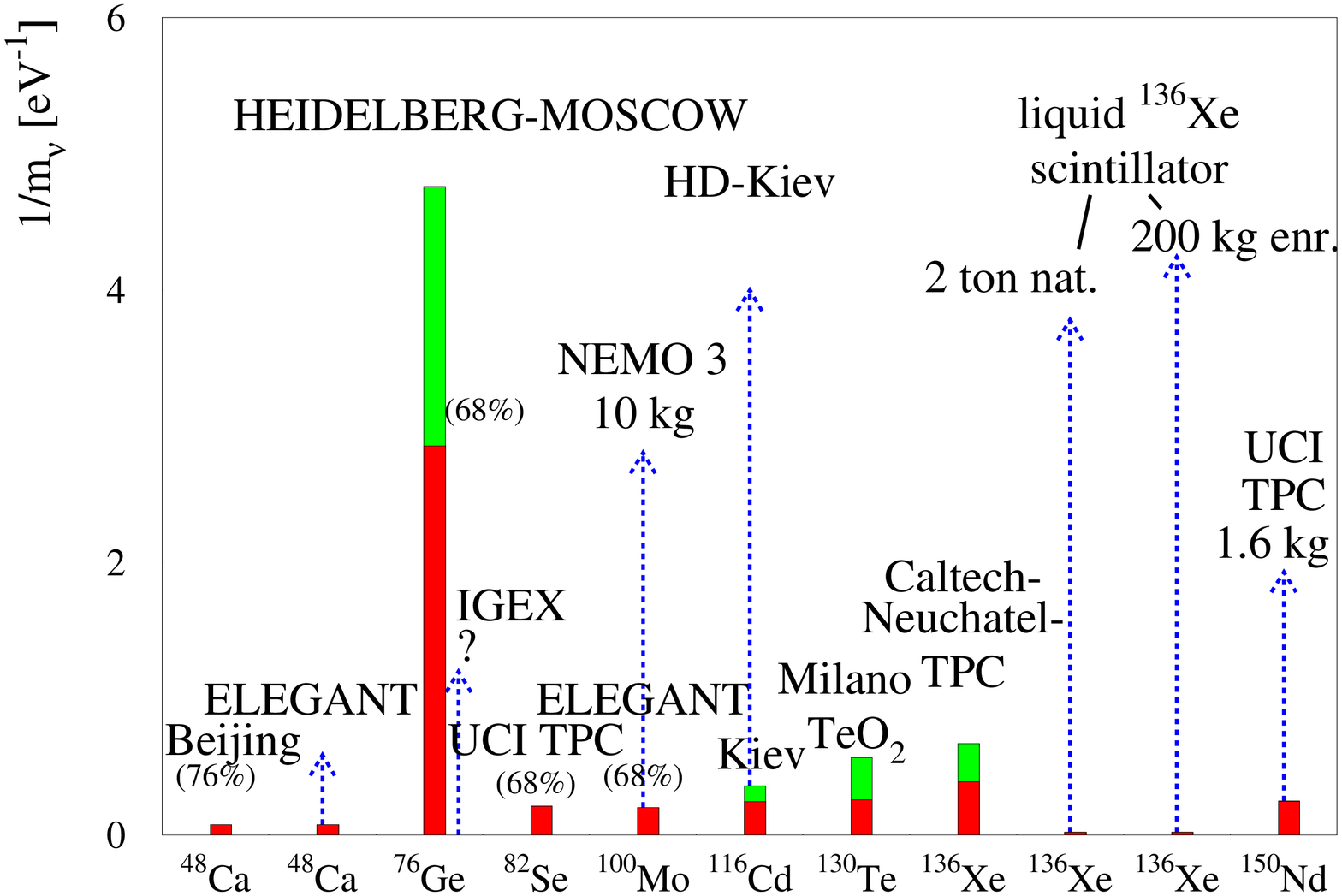,height=6cm,width=9cm}}}
\caption{Sensitivity of ${\beta \beta}_{0\nu}$ experiments. }
\label{betabetafut}
\end{figure}

Present data place an important limit on a weighted average \neu mass
parameter $\VEV{m} \lsim 1 - 2$ eV. The present experimental situation
as well as future prospects is illustrated in \fig{betabetafut}, taken from
ref. \cite{Klapdor}.  
Note that this bound depends to some extent on the relevant nuclear
matrix elements characterising this process \cite{haxtongranada}.  The
parameter $\VEV{m}$ involves both neutrino masses and mixings. Thus,
although rather stringent, this limit may allow very large \neu
masses, as there may be strong cancellations between different
neutrino types. This may happen automatically in the presence of
suitable symmetries. For example, the decay vanishes if the
intermediate neutrinos are Dirac-type, as a result of the
corresponding lepton number symmetry \cite{QDN}.

Neutrino-less double beta decay has a great conceptual importance. It
has been shown \cite{BOX} that in a gauge theory of the weak
interactions a non-vanishing ${\beta \beta}_{0\nu}$ decay rate
requires \neus to be Majorana particles, {\sl irrespective of which
mechanism} induces it. This is important since in a gauge theory
neutrino-less double beta decay may be induced in other ways,
e.g. via scalar boson exchange.

\subsection{Limits from Cosmology }
\vskip .1cm

There are a variety of cosmological arguments that give 
information on neutrino parameters. In what follows I briefly 
consider the critical density and the primordial Nucleosynthesis 
arguments.

\subsubsection{The Cosmological Density Limit }

The oldest cosmological bound on neutrino masses follows from avoiding
the overabundance of relic neutrinos \cite{KT} 
\beq 
\label{RHO1}
\sum m_{\nu_i} \lsim 92 \: \Omega_{\nu} h^2 \: eV\:, 
\eeq 
where $\Omega_{\nu} h^2 \leq 1$ and the sum runs over all species of
isodoublet neutrinos with mass less than $O(1 \: MeV)$. Here
$\Omega_{\nu}=\rho_{\nu}/\rho_c$, where $\rho_{\nu}$ is the neutrino
contribution to the total density and $\rho_c$ is the critical
density.  The factor $h^2$ measures the uncertainty in the present
value of the Hubble parameter, $0.4 \leq h \leq 1$, and $\Omega_{\nu}
h^2$ is smaller than 1.  For the $\nu_{\mu}$ and $\nu_{\tau}$ this
bound is much more stringent than the laboratory limits \eq{1}.

Apart from the experimental interest \cite{jj}, an MeV tau neutrino
also seems interesting from the point of view of structure formation
\cite{ma1}. Moreover, it is theoretically viable as the constraint in
\eq{RHO1} holds only if \neus are stable on the relevant cosmological
time scale. In models with spontaneous violation of total lepton
number \cite{CMP} there are new interactions of neutrinos with the
majorons which may cause neutrinos to decay into a lighter \neu plus a
majoron, for example \cite{fae},
\beq
\label{NUJ}
\nu_\tau \ra \nu_\mu + J \:\: .
\eeq
or have sizeable annihilations to these majorons,
\beq
\label{nunuJJ}
\nu_\tau + \nu_\tau \ra J + J \:\: .
\eeq

The possible existence of fast decay and/or annihilation channels
could eliminate relic neutrinos and therefore allow them to have
higher masses, as long as the lifetime is short enough to allow for an
adequate red-shift of the heavy neutrino decay products. These 2-body
decays can be much faster than the visible modes, such as radiative
decays of the type $\nu' \ra \nu + \gamma$. Moreover, the majoron
decays are almost unconstrained by astrophysics and cosmology (for 
a detailed discussion see ref. \cite{KT}).

A general method to determine the majoron emission decay rates of
neutrinos was first given in ref. \cite{774}. The resulting decay
rates are rather model-dependent and will not be discussed here.
Explicit neutrino decay lifetime estimates are given in
ref. \cite{Romao92,fae,V}.  The conclusion is that there are many ways
to make neutrinos sufficiently short-lived and that all mass values
consistent with laboratory experiments are cosmologically acceptable.

\subsubsection{The Nucleosynthesis Limit}

There are stronger limits on neutrino lifetimes or annihilation cross
sections arising from cosmological nucleosynthesis. Recent data on the
primordial deuterium abundance \cite{dmeas} have stimulated a lot
of work on the subject \cite{cris.ncris}.  If a massive \nt is
stable on the nucleosynthesis time scale, ($\nu_\tau$ lifetime longer
than $\sim 100$ sec), it can lead to an excessive amount of primordial
helium due to their large contribution to the total energy
density. This bound can be expressed through an effective number of
massless neutrino species ($N_\nu$). If $N_\nu < 3.4-3.6$, one can
rule out $\nu_\tau$ masses above 0.5 MeV \cite{KTCS91,DI93}.  If we
take $N_\nu < 4$ the \mnt limit loosens accordingly. However it has
recently been argued that non-equilibrium effects from the light
neutrinos arising from the annihilations of the heavy \nt's make the
constraint a bit stronger in the large \mnt region \cite{noneq}. In
practice, all $\nu_\tau$ masses on the few MeV range are ruled out.
One can show, however that in the presence of \nt annihilations the
nucleosynthesis \mnt bound is substantially weakened or eliminated
\cite{DPRV}.  Fig. 4 gives the effective number of massless neutrinos
equivalent to the contribution of a massive \nt majoron model with
different values of the coupling $g$ between $\nu_\tau$'s and $J$'s,
expressed in units of $10^{-5}$. For comparison, the dashed line
corresponds to the SM $g=0$ case. One sees that for a fixed
$N_\nu^{max}$, a wide range of tau neutrino masses is allowed for
large enough values of $g$. No \nt masses below the LEP limit can be
ruled out, as long as $g$ exceeds a few times $10^{-4}$.
\begin{figure}[t]
\centerline{\protect\hbox{\psfig{file=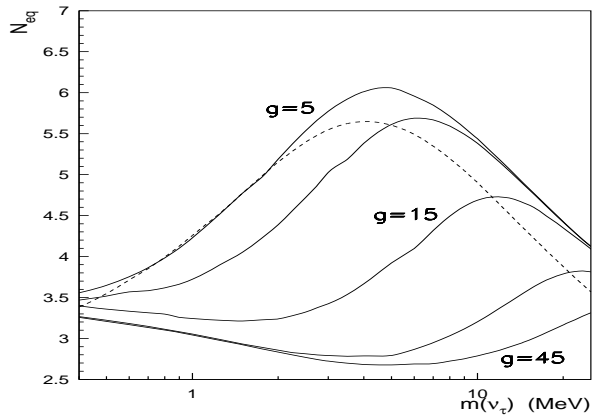,height=5.5cm,width=8cm}}}
\caption{A heavy \nt annihilating to majorons can lower
the equivalent massless-neutrino number in nucleosynthesis.}  
\label{neq} 
\end{figure} 
One can express the above results in the $m_{\nu_\tau}-g$ plane, as
shown in figure \ref{neffmg}.  
\begin{figure}[t]
\centerline{
\psfig{file=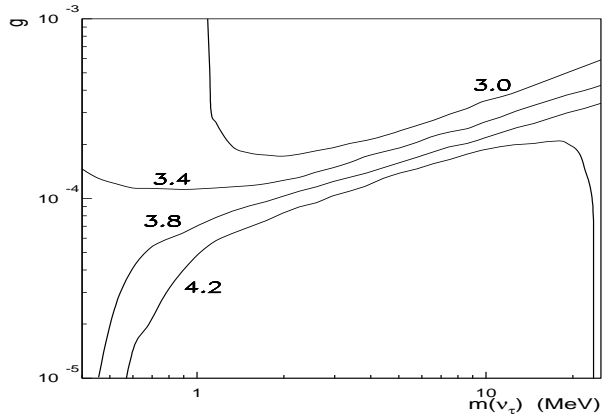,height=5.5cm,width=8cm}}
\caption{The region above each curve is allowed for the
corresponding $N_{eq}^{max}$.}  
\vglue -.5cm 
\label{neffmg}
\end{figure} 
One sees that the constraints on the mass of a Majorana $\nu_\tau$
from primordial nucleosynthesis can be substantially relaxed if
annihilations $\nu_\tau \bar{\nu}_\tau \leftrightarrow JJ$ are
present. Moreover the required values of $g(m_{\nu_\tau})$ are
reasonable in many majoron models \cite{fae,DPRV,MASIpot3}.  Similar
depletion in massive \nt relic abundance also happens if the \nt is
unstable on the nucleosynthesis time scale \cite{unstable} as will
happen in many majoron models.

\subsection{Limits from Astrophysics  }
\vskip .1cm

There are a variety of limits on neutrino parameters that follow from
astrophysics, e.g. from the SN1987A observations, as well as
from supernova theory, including supernova dynamics \cite{Raffelt} and
from nucleosynthesis in supernovae \cite{qian}. Here I briefly discuss
three recent examples of how supernova physics constrains neutrino
parameters.

\begin{figure}[t]
\centerline{\protect\hbox{
\psfig{file=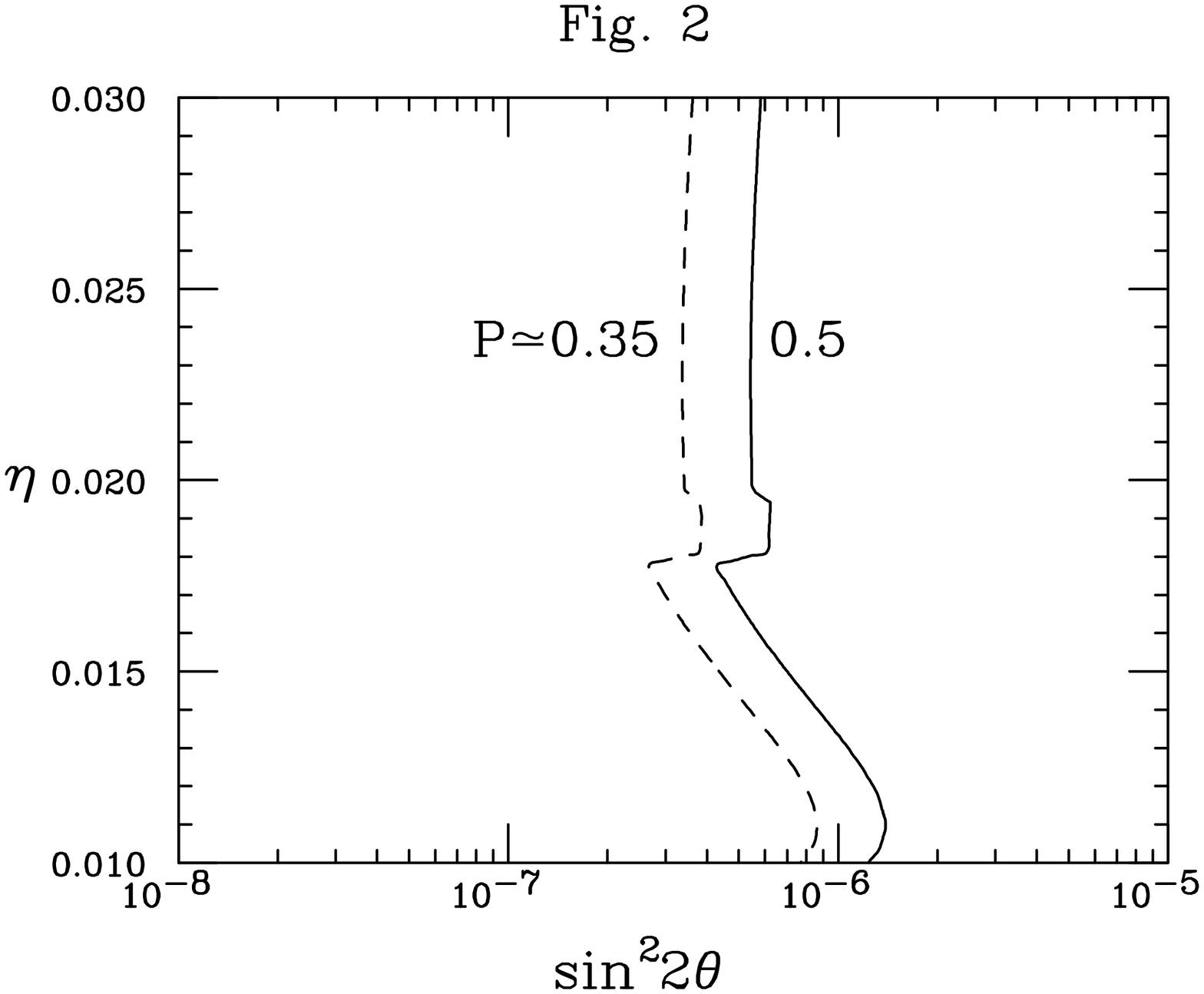,height=6cm,width=8cm,angle=90}
}}
\vglue -6.2cm
\hglue 7.5cm
\psfig{file=cover2.ps,height=.7cm,width=3cm,angle=90}
\vglue 5cm
\caption{SN1987A bounds on massless neutrino mixing. }
\label{SN87}
\end{figure}

It has been noted a long time ago that, in some circumstances, {\sl
massless} neutrinos may be {\sl mixed} in the leptonic charged current
\cite{massless0}. Conventional neutrino oscillation searches in vacuo
are insensitive to this mixing. However,  such neutrinos may
resonantly convert in the dense medium of a supernova
\cite{massless0,massless}. The observation of the energy spectrum of
the SN1987A $\bar{\nu}_e$'s \cite{ssb} may be used to provide very
stringent constraints on {\sl massless} neutrino mixing angles, as
seen in \fig{SN87}.  The regions to the right of the solid curves are
forbidden, those to the left are allowed. Massless neutrino mixing may
also have important implications for $r$-process nucleosynthesis in
the supernova \cite{qian}. For details see ref. \cite{massless}.

\begin{figure}[t]
\centerline{\protect\hbox{
\psfig{file=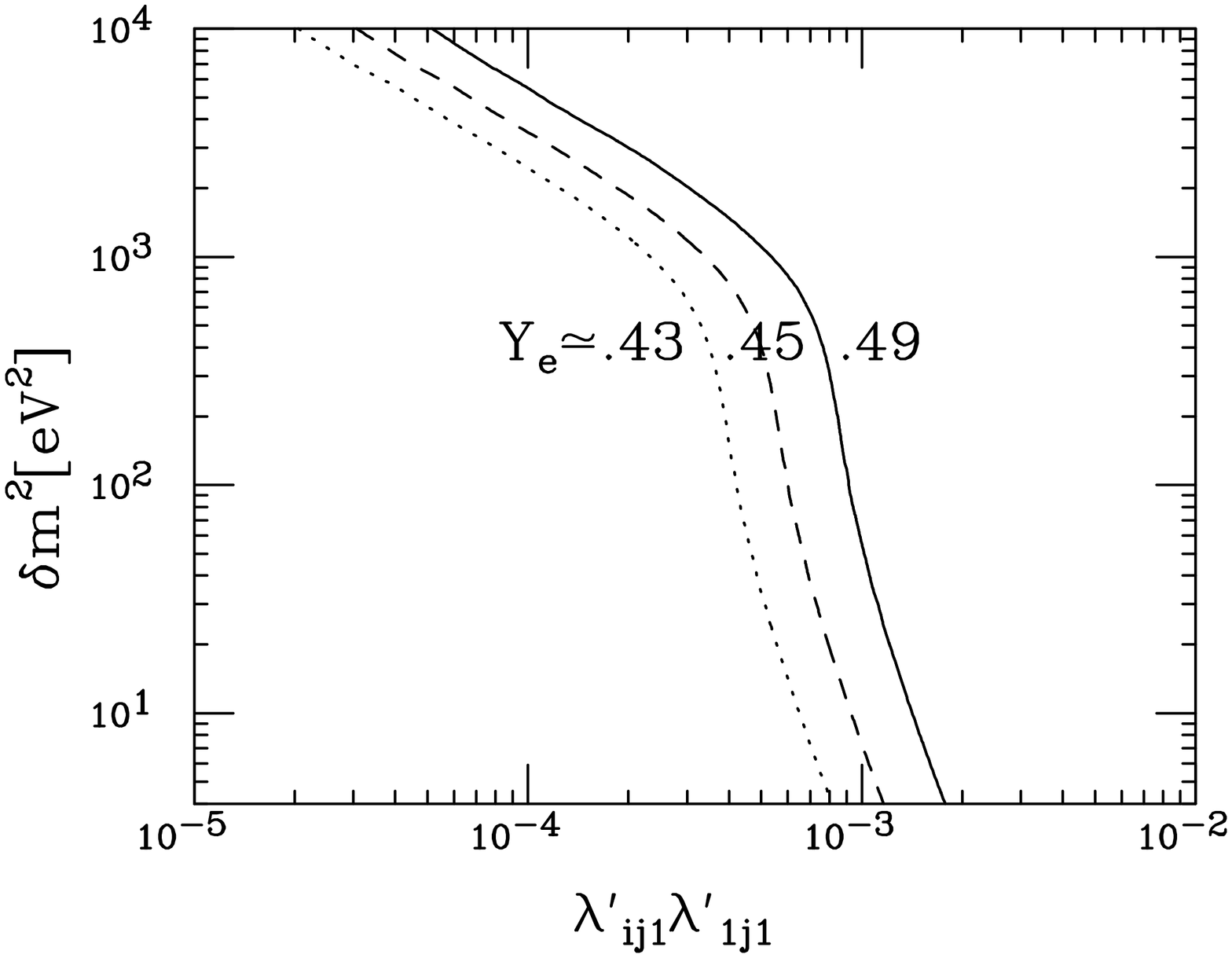,height=6cm,width=8cm,angle=90}
}}
\vglue -.7cm
\caption{Supernovae and FCNC neutrino interactions. }
\label{fcncye2}
\end{figure}

\begin{figure}[t]
\centerline{\protect\hbox{
\psfig{file=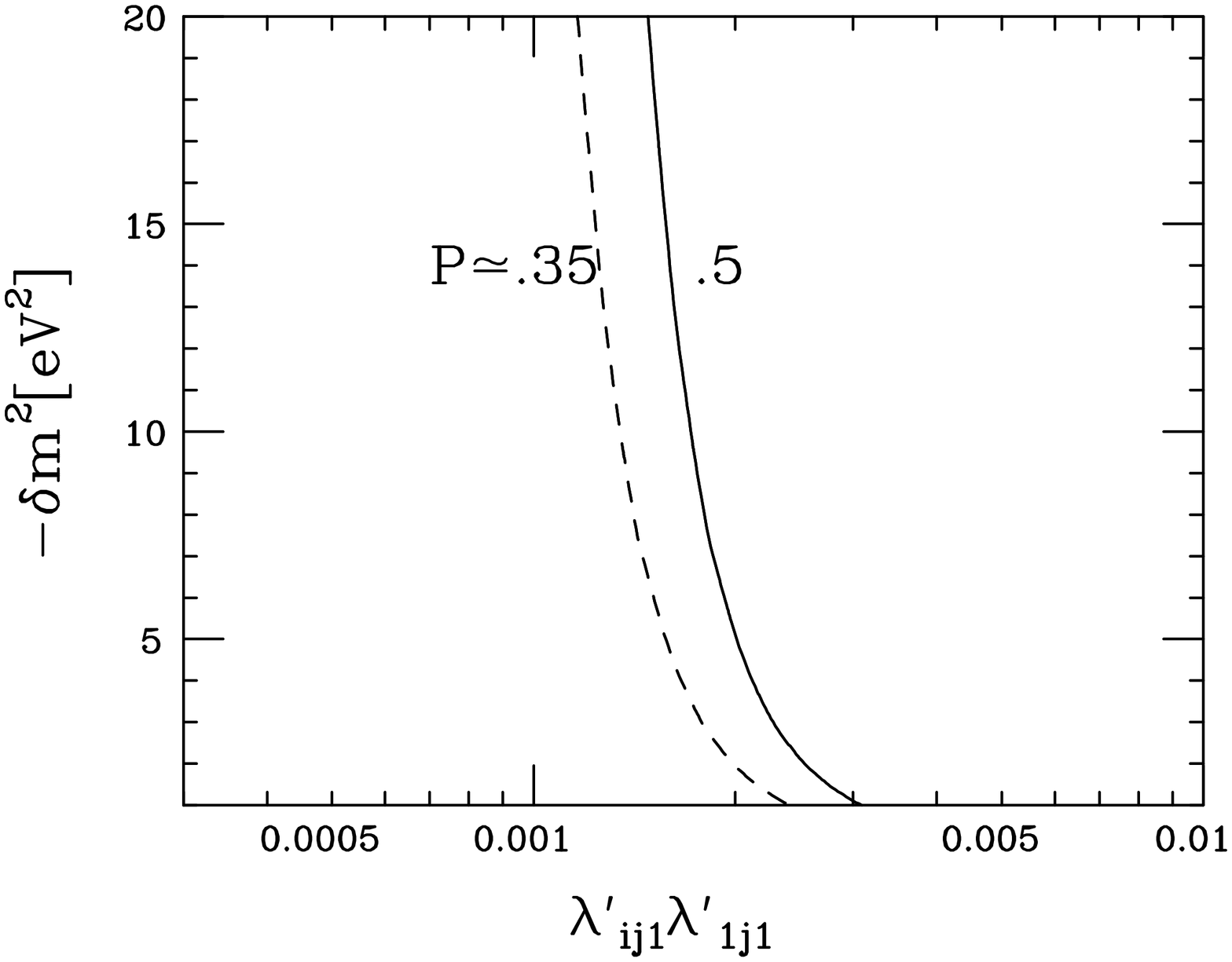,height=6cm,width=8cm,angle=90}
}}
\vglue -.7cm
\caption{Supernovae and FCNC neutrino interactions. }
\label{fcncprob2}
\end{figure}

Another illustration of how supernova restricts \neu properties has
been recently considered in ref. \cite{rsusysn}. There flavour
changing neutral current (FCNC) \neu interactions were considered.
These may induce resonant massless-neutrino conversions in a dense
supernova medium, both in the massless and massive case. The
restrictions that follow from the observed $\bar\nu_e$ energy spectra
from SN1987A and the supernova $r$-process nucleosynthesis provide
constraints on supersymmetric models with $R$ parity violation, which
are much more stringent than those obtained from the laboratory. In
\fig{fcncye2} and \fig{fcncprob2} we display the constraints on explicit
$R$-parity-violating FCNCs in the presence of non-zero neutrino masses
in the hot dark matter eV range.  As seen from \fig{fcncprob2} and
\fig{fcncye2} they isfavour a leptoquark interpretation of the 
recent HERA anomaly.

\begin{figure}[t]
\centerline{\protect\hbox{\psfig{file=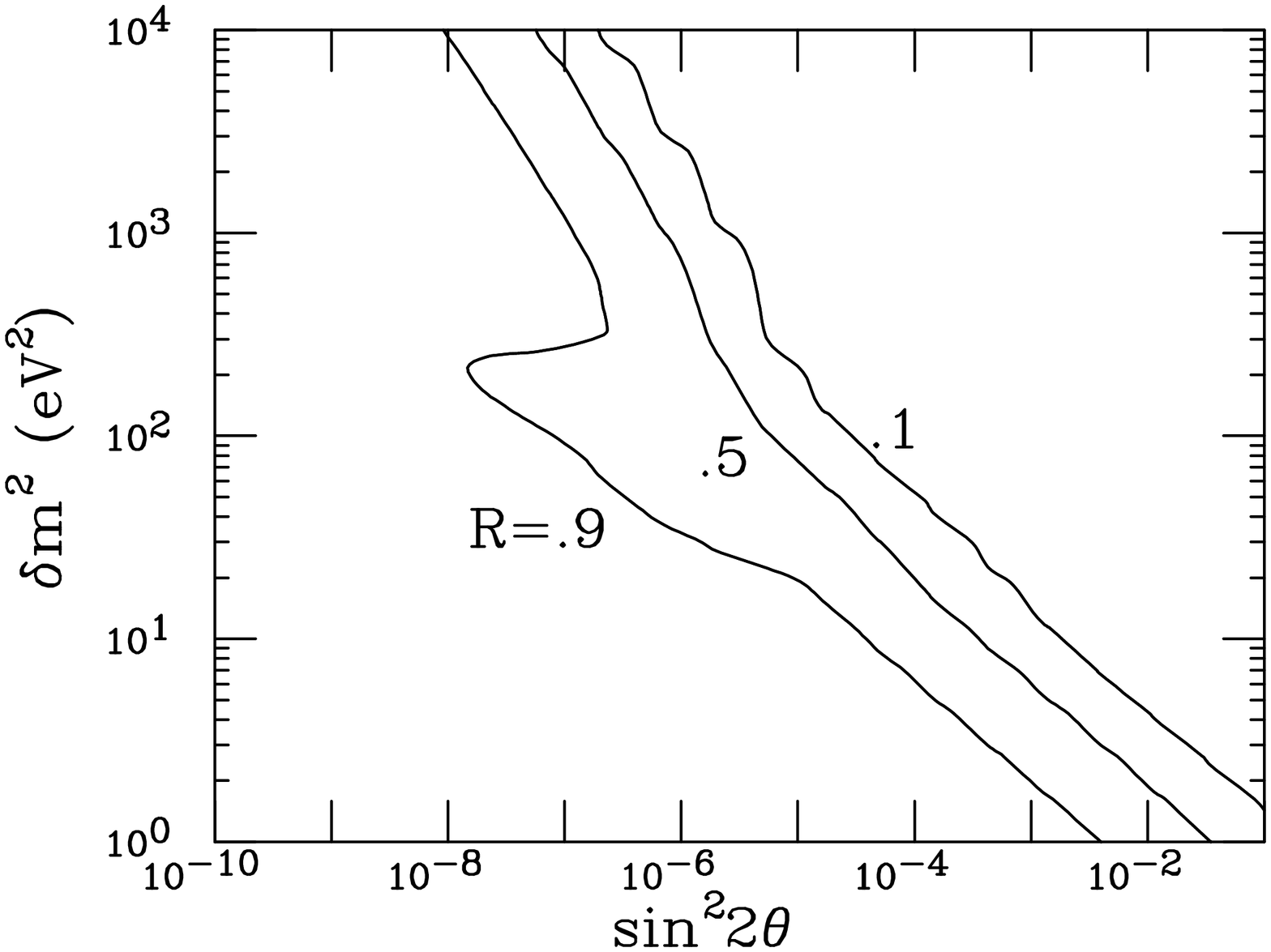,height=6cm,width=7cm,angle=90}}}
\vglue -0.3cm
\caption{Supernovae and sterile neutrinos. }
\label{sterileSN}
\end{figure}

As a final example of how astrophysics can constrain \neu properties
we consider the case of resonant $\nu_e \to\nu_s$ and
$\bar{\nu}_e\to\bar{\nu}_s$ conversions in supernovae, where
$\nu_s$ is a {\it sterile} neutrino \cite{nunus}, which we assume to
be in the hot dark matter mass range.  The implications of such a
scenario for the supernova shock re-heating, the detected $\bar\nu_e$
signal from SN1987A and for the $r$-process nucleosynthesis hypothesis
have been recently analysed \cite{nunus}. In \fig{sterileSN}, taken
from \cite{nunus}, we illustrate the resulting constraints on mixing
and mass difference for the $\nu_e-\nu_s$ system that follow from the
supernova shock re-heating argument.  
Notice that for the case of $r$-process nucleosynthesis there is an
allowed region for which the $r$-process nucleosynthesis can be
enhanced.

\section{Indications for Neutrino Mass}
\vskip .1cm

So far most of positive hints in favour of nonzero neutrino rest
masses come from astrophysics and cosmology, with a varying degree of
theoretical assumptions.  We now turn to these.

\subsection{Dark Matter}
\vskip .1cm

Considerations based on structure formation in the Universe have
become a popular way to argue in favour of the need of a massive
neutrino \cite{cobe2}. Indeed, by combining the observations of cosmic
background temperature anisotropies on large scales performed by the
COBE satellite \cite{cobe} with cluster-cluster correlation data
e.g. from IRAS \cite{iras} one finds that it is not possible to fit
well the data on all scales within the framework of the simplest cold
dark matter (CDM) model. The simplest way to obtain a good fit is to
postulate that there is a mixture of cold and hot components,
consisting of about 80 \% CDM with about 20 \% {\sl hot dark matter}
(HDM) and a small amount in baryons.  The best candidate for the hot
dark matter component is a massive neutrino of about 5 eV.  It has
been argued that this could be the tau neutrino, in which case one
might expect the existence of \ne $\ra$ \nt or \nm $\ra$ \nt
oscillations. Searches are now underway at CERN \cite{chorus}, with a
similar proposal at Fermilab. This mass scale is also consistent with
the hints in favour of neutrino oscillations reported by the LSND
experiment \cite{lsnd}.

\subsection{Solar Neutrinos}
\vskip .1cm

\begin{figure}[t]
\centerline{\protect\hbox{
\psfig{file=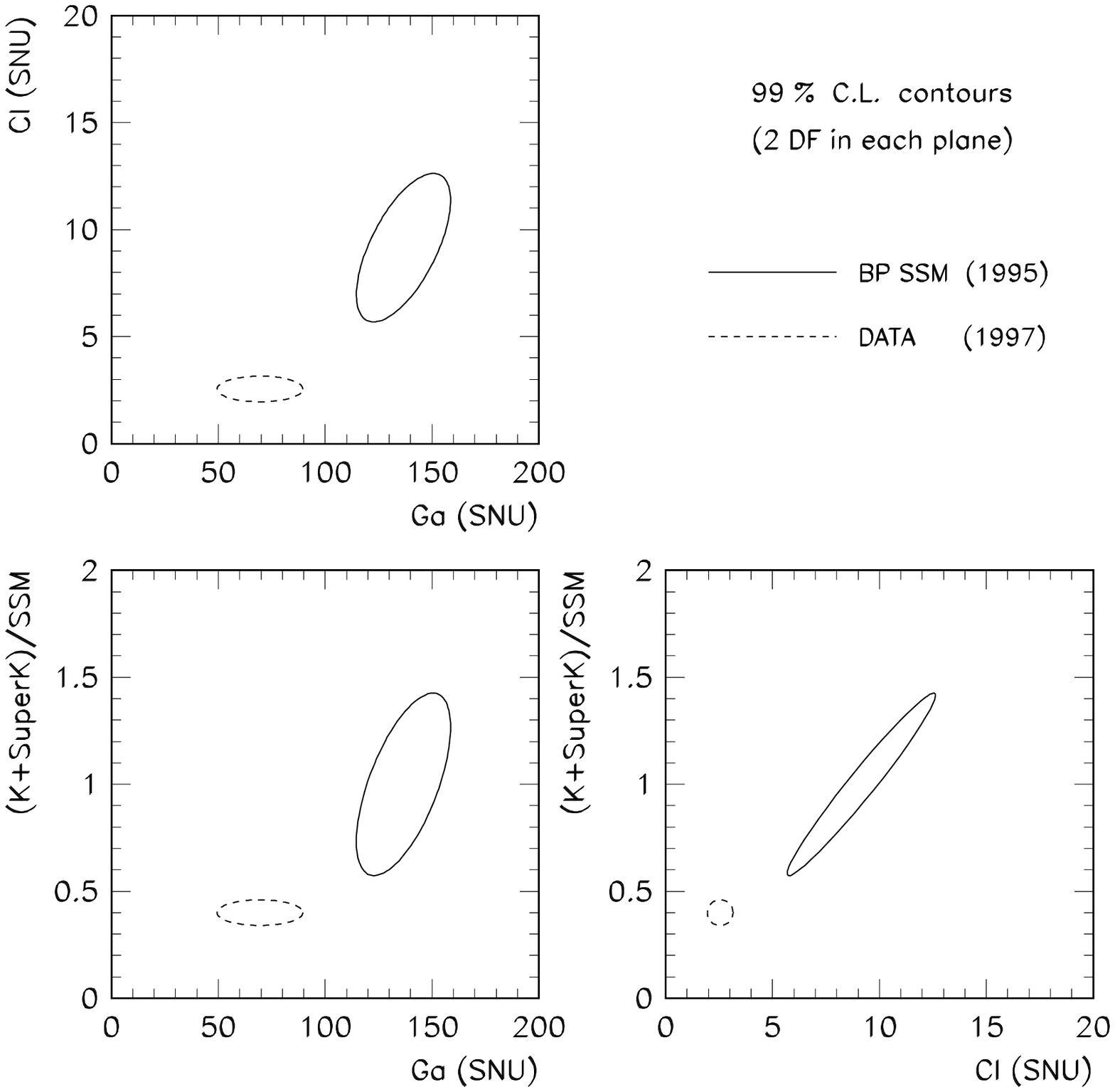,height=11cm,width=11cm}}}
\caption{Solar neutrinos: theory versus data. }
\label{solardata200sk}
\end{figure}

The averaged data collected by the chlorine, Kamiokande, as well as by
the low-energy data on pp neutrinos from the GALLEX and SAGE
experiments still pose a persisting puzzle, now
re-confirmed by the first 200 days of Super-Kamiokande (SK) data
\cite{chiaki}. The most recent data can be summarised in \fig{solardata200sk}
where the theoretical predictions refer to the BP95 SSM prediction of
ref. \cite{SSM}.  For the gallium result we have taken the average of
the GALLEX and the SAGE measurements.

\begin{figure}[t]
\centerline{\protect\hbox{\psfig{file=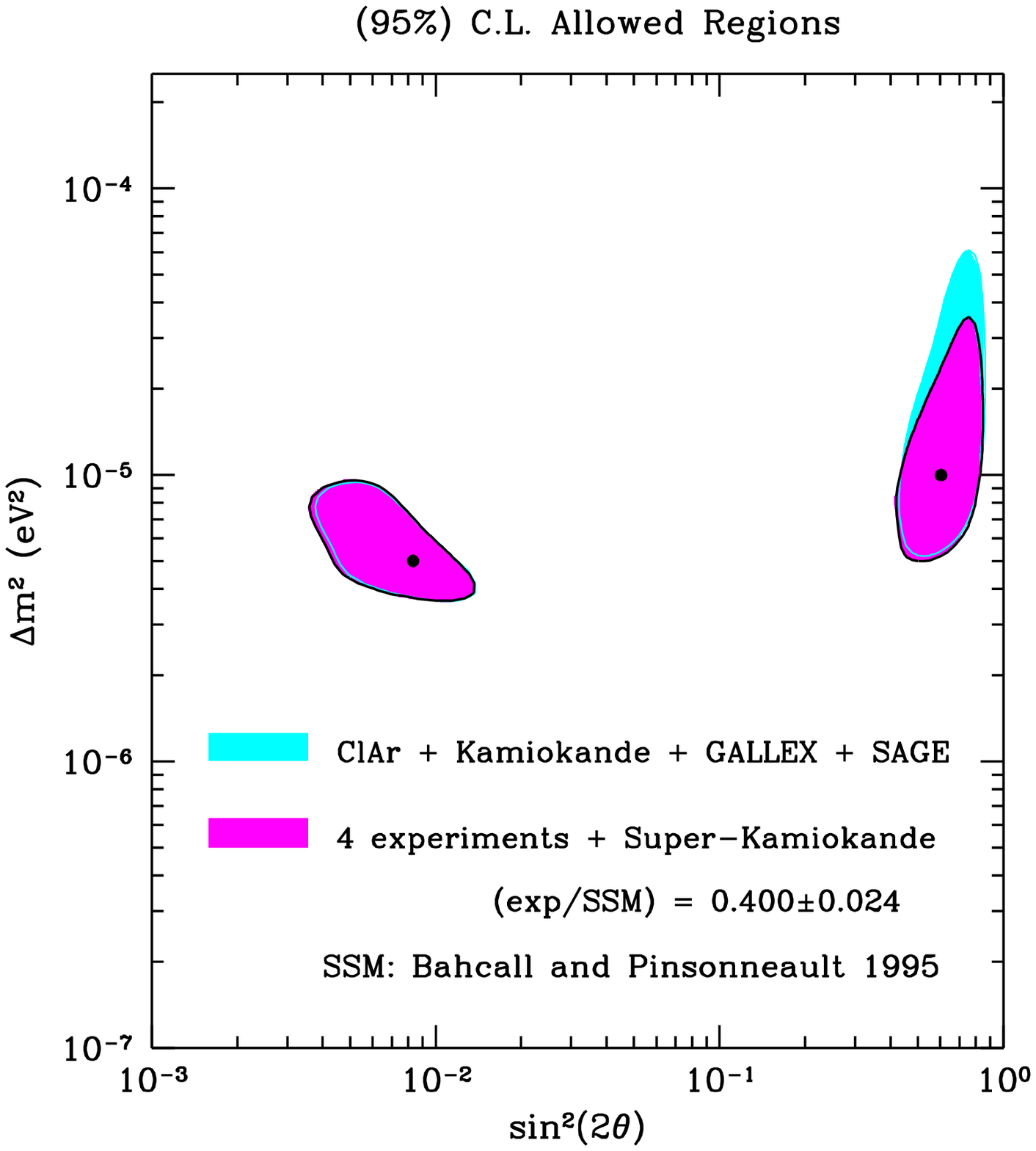,width=8cm,height=7cm}}}
\caption{Solar neutrino parameters for active MSW conversions.}
\label{msw}
\end{figure}

The totality of the data strongly suggests that the solar neutrino
problem is real, that the simplest astrophysical solutions are ruled
out, and therefore that new physics is needed \cite{CF}. The most
attractive possibility is to assume the existence of \neu conversions
involving very small \neu masses. In the framework of the MSW effect
\cite{MSW} the required solar neutrino parameters $\Delta m^2$ and
$\sin^2 2\theta$ are determined through a $\chi^2$ fit of the
experimental data. In \fig{msw} , taken from ref. \cite{bks}, we show the
allowed two-flavour regions obtained in an updated MSW analysis of the
solar \neu data including the the recent SK 200 days data, in the BP95
model for the case of active neutrino conversions. The analysis of
spectral distortion as well as day-night effect plays an important
role in ruling out large region of parameters. Compared with
previously, the impact of the recent SK data is felt mostly in the
large mixing solution which, however, does not give as good a fit as
the small mixing solution, due mostly to the larger reduction of the
$^7$Be flux found in the later. 
The most popular alternative solutions to the solar neutrino anomaly
include the MSW sterile neutrino conversions, as well as the just-so
or vacuum oscillation solution. Recent fits have also been given
including the recent SK data \cite{bks}.

A theoretical point of direct phenomenological interest for Borexino
is the study of the possible effect of random fluctuations in the
solar matter density \cite{BalantekinLoreti}. The existence of noise
fluctuations at a few percent level is not excluded by the SSM nor by
present helioseismology studies. They may strongly affect the $^7$Be
neutrino component of the solar neutrino spectrum so that the Borexino
experiment should provide an ideal test, if sufficiently small errors
can be achieved. The potential of Borexino in "testing" the level of
solar matter density fluctuations is discussed quantitatively in
ref. \cite{noise}.
 
\subsection{Atmospheric Neutrinos}
\vskip .1cm

Water Cerenkov underground experiments, Kamiokande, Superkamiokande
and IMB show a clear deficit in the expected flux of atmospheric
$\nu_\mu$'s relative to that of $\nu_e$'s that would be produced from
conventional decays of $\pi$'s, $K$'s as well as secondary muon decays
\cite{chiaki,Barish}. This is also seen in the iron calorimeter
Soudan2 experiment.  Although the predicted absolute fluxes of \neus
produced by cosmic-ray interactions in the atmosphere are uncertain at
the 20\% level, their ratios are expected to be accurate to within
5\%. Other experiments, such as Frejus and NUSEX, have not found a
firm evidence, but they have much larger errors. It is tempting
therefore to take seriously the evidence for an atmospheric neutrino
deficit and to ascribe to \neu oscillations.  In ref. \cite{atm} the
impact of recent experimental results on atmospheric neutrinos from
experiments such as Superkamiokande and Soudan on the determinations
of atmospheric neutrino oscillation parameters is considered for the
$\nu_\mu \to \nu_\tau$ channel.  In performing this re-analysis
theoretical improvements in flux calculations as well as
neutrino-nucleon cross sections have been taken into account.  The
relevant allowed regions of oscillation parameters have been
determined from a fit of the various data.  One of the new features
that arises from the inclusion of the Superkamiokande data is that the
best fit value of the $\Delta m^2$ is lower than previously obtained.
In fact for the $\nu_\mu \to \nu_e$ channel the allowed region is
almost totally ruled out by the recent Chooz data \cite{Chooz}.

\section{Reconciling Present Hints}
\vskip .1cm

\subsection{Almost Degenerate Neutrinos}
\vskip .1cm

The above observations from cosmology and astrophysics do seem to
suggest a theoretical puzzle. As can easily be understood just on the
basis of numerology, it seems rather difficult to reconcile the three
observations discussed above in a framework containing just the three
known \neus. The only possibility to fit these observations in a world
with just the three known neutrinos is if all of them have nearly the
same mass $\sim$ 2 eV \cite{caldwell}. This can be arranged, for
example in general seesaw models which also contain an effective
triplet VEV \cite{LR,2227} contributing to the light neutrino
masses. This term should be added to \eq{SEESAW}.  Thus one can
construct extended seesaw models where the main contribution to the
light \neu masses ($\sim$ 2 eV) is universal, due to a suitable
horizontal symmetry, while the splittings between \ne and \nm explain
the solar \neu deficit and that between \nm and \nt explain the
atmospheric \neu anomaly \cite{DEG}.

\subsection{Four-Neutrino Models}
\vskip .1cm

A simpler alternative way to fit all the data is to add a fourth \neu
species which, from the LEP data on the invisible Z width, we know
must be of the sterile type, call it \ns. The first scheme of this
type gives mass to only one of the three neutrinos at the tree level,
keeping the other two massless \cite{OLD}.

Two basic schemes of this type that keep the sterile neutrino light
due to a special symmetry have been suggested. In addition to the
sterile \neu \ns, they invoke additional Higgs bosons beyond that of
the SM, in order to generate radiatively the scales
required for the solar and atmospheric \neu conversions. In these
models the \ns either lies at the dark matter scale \cite{DARK92} as
illustrated in \fig{pv}
\begin{figure}[t]
\centerline{\protect\hbox{\psfig{file=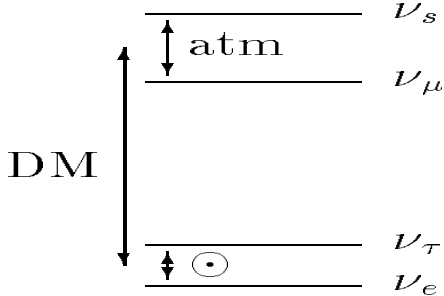,width=6cm,height=4cm}}}
\caption{{\sl "Heavy"} Sterile 4-Neutrino Model}
\label{ptv}
\vglue -.3cm
\end{figure}
or, alternatively, at the solar \neu scale \cite{DARK92B}. 
\begin{figure}
\centerline{\protect\hbox{\psfig{file=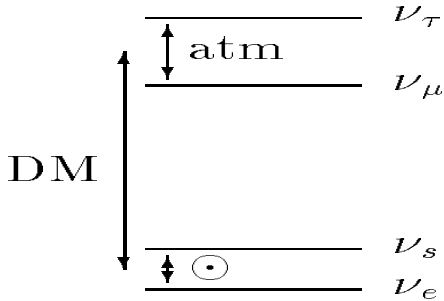,width=6cm,height=4cm}}}
\caption{{\sl Light} Sterile 4-Neutrino Model}
\label{pv}
\vglue -.3cm
\end{figure}
In the first case the atmospheric \neu puzzle is explained by \nm to
\ns oscillations, while in the second it is explained by \nm to \nt
oscillations. Correspondingly, the deficit of solar \neus is explained
in the first case by \ne to \nt oscillations, while in the second it
is explained by \ne to \ns oscillations. In both cases it is possible
to fit all observations together.  However, in the first case there is
a clash with the bounds from big-bang nucleosynthesis. In the latter
case the \ns is at the MSW scale so that nucleosynthesis limits are
satisfied. They nicely agree with the best fit points of the
atmospheric neutrino parameters from Kamiokande \cite{atm}. Moreover,
it can naturally fit the hints of neutrino oscillations of the LSND
experiment \cite{lsnd}.  Another theoretical possibility is that all
active \neus are very light, while the sterile \neu \ns is the single
\neu responsible for the dark matter \cite{DARK92D}.

\subsection{Mev Tau Neutrino}
\vskip .1cm

An MeV range tau neutrino is an interesting possibility to consider
for two reasons. First, such mass is within the range of the
detectability, for example at a tau-charm factory \cite{jj}. On the
other hand, if such neutrino decays before the matter dominance epoch,
its decay products would add energy to the radiation, thereby delaying
the time at which the matter and radiation contributions to the energy
density of the universe become equal. Such delay would allow one to
reduce the density fluctuations on the smaller scales purely within
the standard cold dark matter scenario, and could thus reconcile the
large scale fluctuations observed by COBE \cite{cobe} with the
observations such as those of IRAS \cite{iras} on the fluctuations on
smaller scales.

In ref.~\cite{JV95} a model was presented where an unstable MeV
Majorana tau \neu naturally reconciles the cosmological observations
of large and small-scale density fluctuations with the cold dark
matter model (CDM) and, simultaneously, with the data on solar and
atmospheric neutrinos discussed above. The solar \neu deficit is
explained through long wavelength, so-called {\sl just-so}
oscillations involving conversions of \ne into both \nm and a sterile
species \ns, while the atmospheric \neu data are explained through \nm
$\ra$ \ne conversions. Future long baseline \neu oscillation
experiments, as well as some reactor experiments will test this
hypothesis. The model assumes the spontaneous violation of a global
lepton number symmetry at the weak scale.  The breaking of this
symmetry generates the cosmologically required decay of the \nt with
lifetime $\tau_{\nu_\tau} \sim 10^2 - 10^4$ seconds, as well as the
masses and oscillations of the three light \neus \ne, \nm and \ns
required in order to account for the solar and atmospheric \neu
data. One can verify that the big-bang nucleosynthesis constraints
\cite{KTCS91,DI93} can be satisfied in this model.

\section{Electroweak Symmetry Breaking}
\vskip .1cm

A basic ingredient in the SM is the breaking of the electroweak
symmetry via the Higgs mechanism. If indeed the Higgs boson exists as
an elementary particle, apart from its direct search, the main task in
these investigations is the study of supersymmetric extensions of the
SM and the corresponding experimental searches at high
energy accelerators.

\subsection{ Supersymmetry and the MSSM}
\vskip .1cm

The main theoretical motivation for SUSY are that it allows for a
stable hierarchy between the electroweak scale responsible for the W
and Z masses and the mass scale of unification. With this requirement
it follows that SUSY should be broken at the electroweak scale and
therefore SUSY particles are expected to exist at this scale.  With
this input, one obtains that the gauge coupling constants measured at
LEP and other experiments, when evolved via the renormalization group
equations to high energies, will join at a scale compatible with
proton stability \cite{gunif}.

The simplest SUSY model is the Minimal Supersymmetric Standard Model
(MSSM) \cite{mssm}. This model realizes SUSY in the presence of a
discrete R parity ($R_p$) symmetry. Under this symmetry all standard
model particles are even while their partners are odd. As a result of
this selection rule SUSY particles are only produced in pairs, with
the lightest of them being stable. In the MSSM the Lightest SUSY
Particle, LSP for short, is typically a neutralino, for most choices
of SUSY parameters.  It has been suggested as a candidate for the cold
dark matter of the universe and several methods of detection at
underground installations have been suggested \cite{chicdm}.  However,
one should not forget that R parity is postulated {\sl ad hoc},
without a deep theoretical basis. Moreover there are other ways to
explain the cold dark matter via the axion. Last, but not least, hot
dark matter is needed in any case, not to to mention other existing
puzzles in \neu physics, such as the solar \neu deficit. From this
point of view the emphasis of the simplest MSSM picture would seem
exagerated.

\subsection{Broken R--Parity}
\vskip .1cm

Unfortunately there is no firm theoretical basis as to how SUSY is
realized (if at all) in Nature. Nobody knows the origin of the R
parity symmetry. As a matter of fact it could well be broken via
tri-linear and bi-linear superpotential couplings.  The latter has
been shown to be compatible with minimal supergravity with universal
boundary conditions at unification \cite{epsrad,DJV} as well as the
smallness of \neu masses. This happens for relatively large values of
the relevant model effective R--parity violation parameter $\epsilon$
\footnote{For three generations there are three $\epsilon_i$, but
here for simplicity we focus only on one.}.  It provides the simplest
reference model for the breaking of R--parity, in the same way as the
MSSM provides the simplest phenomenological model for SUSY.

A more satisfactory picture to R--parity violation would be one in
which it is conserved at the Lagrangian level but breaks spontaneously
through a sneutrino VEV.  Keeping the minimal \21 gauge structure this
also implies the spontaneous breaking of lepton number, which is a
continuous ungauged symmetry, and therefore the existence of an
associated Goldstone boson (majoron).  The breaking of R-parity should
be driven by {\sl isosinglet} right-handed sneutrino vacuum
expectation values (VEVS) \cite{MASIpot3} so as to avoid conflicts
with LEP observations of the invisible $Z$ width (in this case the
majoron is mostly singlet, and does not couple appreciably to the Z).
The theoretical viability of this scenario has been demonstrated both
with tree-level breaking of the electroweak symmetry and R--parity
\cite{MASIpot3}, as well as in the most attractive radiative breaking
approach \cite{IRV}.

Typically in these models neutrinos have mass. In the conceptually
simplest models the origin of neutrino mass is the breaking of
R--parity, as in \cite{MASIpot3,IRV,epsrad,DJV,beyond}. In this case the
magnitude of R--parity violating effects is directly correlated with
the neutrino mass. However, other consistent possibilities exist
\cite{sbrpothers}.

The model with bi-linear breaking is specially attractive because of
its simplicity and because it is the effective truncated version of
the more complete models with spontaneous breaking of R--parity.

In the following few sections I will illustrate with some examples the
potential of the present and future colliders in testing supersymmetry
with spontaneous or bilinear breaking of R--parity under the
assumption that neutrinos acquire mass only due to R--parity
violation.  For simplicity we will refer to these models generically
as RPSUSY models. The characteristic feature of these models is
that the pattern of R--parity breaking interactions is determined in
terms of relatively few new parameters in addition to those of the
MSSM (one in the simplest reference model).  This allows for a
systematic discussion of the potential of new colliders in searching
for broken R--parity SUSY signals.

\subsubsection{R--Parity Violation at LEP}

In the MSSM the usual neutralino pair-production process,
\begin{equation}
\label{chichi}
e^+ e^- \to Z \to \chi \chi
\end{equation}
where $\chi$ denotes the lightest neutralino, leads to no
experimentally detectable signature (other than the contribution to
the Z invisible width), as $\chi$ escapes the apparatus without
leaving any tracks. The simplest process that leads to a zen-event
topology, with particles in one hemisphere and nothing on the
opposite, requires the production of $\chi$ associated to
$\chi^\prime$, $Z \to e^+ e^- \to \chi \chi^\prime$, $\chi^\prime$
being the next-to-lightest neutralino.

In broken R--parity models the $\chi$ may decay into charged
particles, so that \eq{chichi} can lead to zen-events in which one
neutralino decays visibly (leptons and jets) and the other
invisibly. The topology is the same as in the MSSM but the
corresponding zen-event rates can be larger than in the MSSM and may
occur below the threshold for $\chi^\prime$ production.  The missing
momentum in these models is carried by the \nt or by majorons.
Another possibility for zen events in RPSUSY is the decay $Z \to \chi
\nu_\tau$. Since the latter violates R--parity, the rates are somewhat
smaller, see. Table 1.

\begin{figure}[t]
\centerline{\protect\hbox{\psfig{file=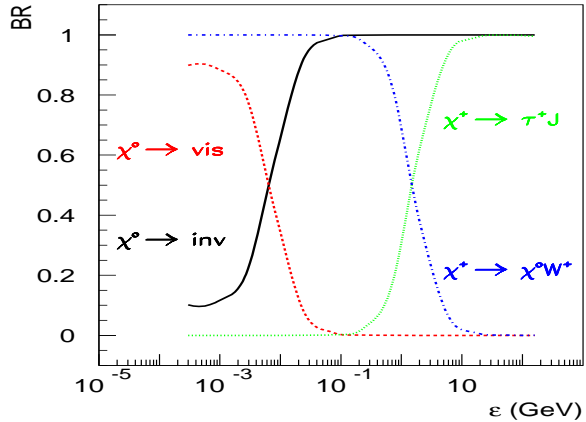,height=6cm,width=9cm}}}
\vglue -0.3cm
\caption{Typical neutralino and chargino decay branching ratios as a
function of $\epsilon $ for $\mu = 150$ GeV, $M_2 = 100$ GeV, and
$\tan\beta = 35$.}
\label{br}
\end{figure}

For the sake of illustration we exhibit in \fig{br} typical values of
the branching ratios of neutralinos and charginos, as a function of
$\epsilon$ for $\mu=150$ GeV, $M_2=100$ GeV, and $\tan\beta=35$.  For
neutralinos we exhibit its total visible and invisible branching
ratios, where we included in the invisible width the contributions
coming from the neutrino plus majoron channel ($\chi \to \nu$J), as
well as from the neutral current channel when the $Z$ decays into a
pair of neutrinos ($\chi \to 3 \nu$).

\begin{figure}[t]
\centerline{\protect\hbox{\psfig{file=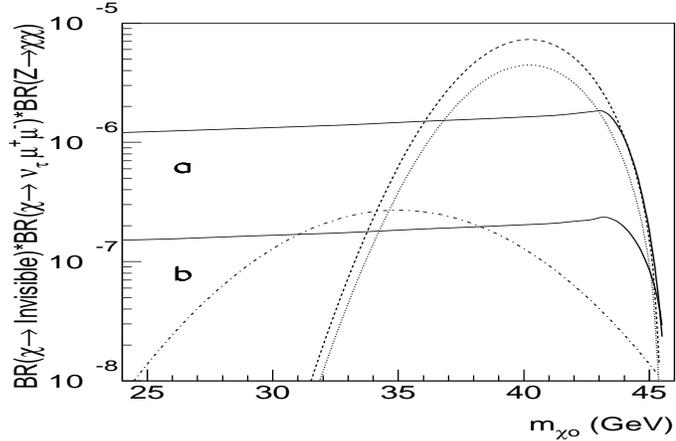,height=6cm,width=9cm}}}
\vglue -0.3cm
\caption{Limits on $BR (Z \to \chi \chi) BR(\chi \to \mu^+\mu^- \nu)$ 
compared with the maximum theoretical values expected in different
RPSUSY models}
\label{chichiab}
\end{figure}

In \fig{chichiab} we illustrate the sensitivity of LEP experiments to
leptonic signals associated to neutralino pair-production at the Z
peak in our RPSUSY models. The signal topology used was missing
transverse momentum plus acoplanar muon events ($ p\!\!\!/_T +\mu^+
\mu^- $) arising from $\chi \chi$ production followed by $\chi$
decays.  The solid line (a) in \fig{chichiab} is the region of
sensitivity of LEP I data of ref. \cite{aleph95} corresponding to an
integrated luminosity of 82 $pb^{-1}$, while (b) corresponds to the
improvement expected from including the $e^+e^-\nu$ channel, as well
as the combined statistics of the four LEP experiments.  The dashed
line corresponds to the bi-linear model of explicit R--parity
violation, allowing \mnt values as large as the present limit, the
dotted one does implement the restriction on \mnt suggested by
nucleosynthesis, and the dash-dotted one is calculated in the model
with spontaneous breaking of R--parity (majoron model).  The inclusion
of semi-leptonic decays and of the updated integrated luminosity
already achieved at LEP would substantially improve the
statistics and thus the sensitivity to RPSUSY parameters.

The usual chargino pair-production process,
\begin{equation}
e^+ e^- \to Z \to \chi^+ \chi^-
\end{equation}
may also provide novel signatures which would not be possible in the
MSSM, as the neutralinos produced from chargino decays may themselves
decay into jets or leptons leading to exotic channels. 

Moreover, in \21 models with spontaneous violation of R--parity the
presence of the majoron implies the existence of two--body chargino
decays \cite{ROMA}
\begin{equation}
\label{tj}
\chi^\pm \to \tau^\pm + J
\end{equation}
In ref. \cite{tauj} chargino pair production at LEP II has been
studied in supersymmetric models with spontaneously broken
$R$--parity. Through detailed signal and background analyses, it was
shown that a large region of the parameter space of these models can
be probed through chargino searches at LEP II.
\begin{figure}[t]
\centerline{\protect\hbox{\psfig{file=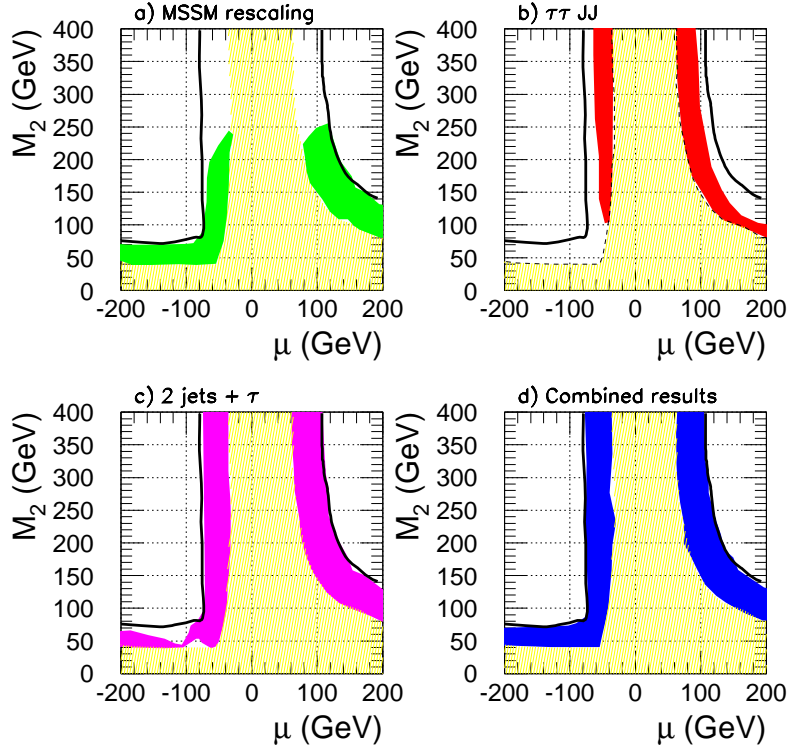,height=12cm,width=12cm}}}
\vglue -0.3cm
\caption{95\% CL excluded region in RPSUSY models in various analyses
(dark areas), as well as the combined excluded region for $\tan
\beta =2$, $\epsilon = 1$ GeV, $\protect\sqrt{s}=172$ GeV, and an
integrated luminosity of 300 pb$^{-1}$.}
\label{zone2-1}
\end{figure}

The limits on the chargino mass depend on the magnitude of the
effective $R$--parity violation parameter $\epsilon$.  As $\epsilon
\to 0$ we recover the usual MSSM chargino mass limits, however, for
$\epsilon$ sufficiently large, the bounds on the chargino mass can be
about 15 GeV weaker than in the MSSM due to the dominance of the
two-body chargino decay mode \eq{tj}. This happens because of the
irreducible background from W-pair production with each $W \to \tau
\nu$. 

Although the \nt can be quite relatively heavy in these models, it is
consistent with the cosmology critical density \cite{KT} as well as
primordial nucleosynthesis \cite{unstable}, due to the existence of the
majoron which opens new \nt decay and annihilation channels
\cite{fae,V}. The small  mass difference between \ne and \nm
may lead to an explanation of solar \neu deficit by resonant \ne to
\nm conversions \cite{MSW}. In this model one may regard the the 
R--parity violating processes as a tool to probe the physics
underlying the solar \neu conversions \cite{RPMSW}. For example, the
rates for some RPSUSY rare decays (see section 5.2.3 below) can be
used in order to discriminate between large and small mixing angle MSW
solutions to the solar \neu problem \cite{MSW}.

\subsubsection{R--Parity Violation at LHC}

\begin{figure}[t]
\centerline{\protect\hbox{\psfig{file=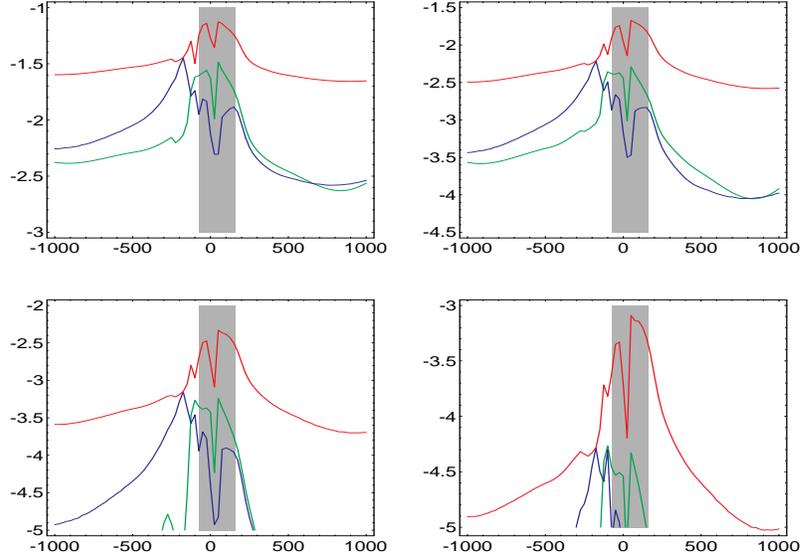,height=11cm,width=11cm}}}
\vglue -0.3cm
\caption{Multi-lepton rates at LHC in various RPSUSY models. }
\label{Clepmlt02}
\end{figure}

It is also possible to find manifestations of R parity violation at
the super-high energies available at hadron super-colliders such as the
Tevatron and the LHC. If SUSY particles, gluinos and squarks, are pair
produced at hadron collisions, their subsequent cascade decays will
not terminate at the lightest neutralino but it will further decay.
To the extent that this decay is into charged leptons it will give
rise to a quite rich pattern of high multiplicity lepton events. Such
pattern of gluino cascade decays in RPSUSY models was studied in
detail in ref. \cite{gluino}. The conclusion is that multi-lepton and
same-sign dilepton signal rates which can be substantially higher than
those predicted in the MSSM. This is illustrated in \fig{Clepmlt02},
which shows the branching ratios for various multi-lepton signals
(summed over electrons and muons) with the 3-, 4-, 5- and 6-leptons,
for $\tan \beta = 2$, with other parameters chosen in a suitable way
(see ref. \cite{gluino} for details).  We show a) the 3-lepton, b) the
4-lepton, c) the 5-lepton and d) the 6-lepton signal for the MSSM
(full line), the majoron-model (dashed line) and the $\epsilon$-model
(dashed-dotted line). The shaded area will be covered by LEP2. Note,
for example, that for $\mu < 0$ the 5-lepton signal is much larger in
the majoron-model than in the MSSM, giving about 30 to 1200 events per
year for an LHC luminosity of $10^{5}pb^{-1}$.  The 6-lepton signal
has a rate up to $5 \times 10^{-5}$ in the range $-300$~GeV$< \mu <
-80$~GeV giving 125 events per year. The multi-lepton rates would be
even higher in the $\epsilon$ model.

Although with smaller rates, one also expects in RPSUSY models the
single production of the SUSY states in hadron collisions, as has been
discussed. In ref. \cite{RPLHC} the single production of weakly
interacting SUSY fermions (charginos and neutralinos) via the
Drell-Yan mechanism was studied.

\subsubsection{Rare Decays}

If R--parity is broken spontaneously it shows up in the couplings of
the W and the Z. As a result there may be rare $Z$ decays with single
production of the charginos \cite{ROMA}, 
\begin{equation}
Z \to \chi^{\pm} \tau^{\mp}
\end{equation}
As mentioned in the RPSUSY models, the magnitude of R parity violation
is correlated with the nonzero value of the \nt mass and is restricted
by a variety of experiments. Nevertheless the R parity violating Z
decay branching ratios, as an example, can easily exceed $10^{-5}$,
well within present LEP sensitivities.

Similarly, the lightest neutralino (LSP) could also be singly-produced
as \cite{ROMA}
\beq
Z \to \chi \nu_\tau
\eeq
In models with spontaneous lepton number violation, like majoron
models of neutrino mass \cite{monorindani}, and the SUSY model with
spontaneous violation of R--parity \cite{monorp}, there are Z decay
processes with single photon emission
\bea
\label{1gamma}
Z \to \gamma + H \\\nonumber
Z \to \gamma + J \\\nonumber
Z \to \gamma + J + J 
\eea
where H is a CP-even Higgs boson and J denotes the associated CP-odd
majoron. Since lepton number violation occurs in these models at the
weak scale, these processes may have relatively high rates, as seen in
Table 1. Their existence would give rise to some rare Z decay
signatures which could potentially be observable at the Z peak.  In
the RPSUSY majoron model \cite{monorp} the first two processes in
\eq{1gamma} violate R--parity and are therefore strictly correlated
with the \nt mass. On the other hand in some majoron masses with
radiatively induced \neu mass, like in \fig{2loop}, the branching
ratios can be relatively large (possibly accessible at LEP), even
though the loop-induced neutrino masses are very small, as required in
order to explain the deficit of solar neutrinos.

Another possible signal of the RPSUSY models based on the simplest \21
gauge group is rare decays of muons and taus such as \cite{NPBTAU}
\bea
\label{mutau}
\tau \to \mu + J \\\nonumber
\tau \to e + J \\\nonumber
\mu \to e + J
\eea
Such decays would be "seen" as bumps in the final lepton energy
spectrum, at half of the parent lepton mass in its rest frame.  Again,
since in this model the lepton number is broken close to the weak
scale it can lead to relatively large rates for single majoron
emitting $\mu$ and $\tau$ decays \cite{NPBTAU} compatible with present
sensitivities and quite interesting for future tau-charm and B
factories \cite{TTTAU}.

Table \ref{rare} summarizes the expectations for rare decay branching
ratios in the class of models of interest. It is interesting to note
that in the broken R--parity majoron since the processes in \eq{mutau}
violate R--parity, their branching ratios are correlated with the \nt
mass. However it was shown in ref.  \cite{NPBTAU} that they can be
large for moderately small \nt masses. This illustrates again how the
search for rare decays can be a more sensitive probe of \neu
properties than the more direct searches for \neu masses, and
therefore complementary.
\begin{table}
\begin{center}
\caption{Allowed rare decay branching ratios in the models considered
in the text.  Here $\chi$ denotes the lightest chargino, $\chi$ the
lightest neutralino, and J is the majoron.}
\begin{displaymath}
\begin{array}{|c|cr|} 
\hline
\mbox{channel} & \mbox{strength} & \mbox{} \\
\hline
Z \to \chi + \tau &  \lsim \mbox{few} \times 10^{-5} & \\
Z \to \chi + \nt &  \lsim 10^{-4} & \\
\hline
Z \to \gamma + J \mbox{ and} Z \to \gamma + H &  \lsim 10^{-6} & \\
Z \to \gamma + J + J &  \lsim 10^{-6} & \\
\hline
\tau \to \mu + J &  \lsim 10^{-3} & \\
\tau \to e + J &  \lsim 10^{-4} & \\
\hline
\end{array}
\end{displaymath}
\end{center}
\label{rare}
\end{table}

\subsection{Scalar Sector}
\vskip .1cm

Although quite indirect, another possible manifestation of the
properties of \neus is in the electroweak breaking sector. Many
extensions of the lepton sector seek to give masses to \neus through
the spontaneous violation of an ungauged U(1) lepton number symmetry,
thus implying the existence of a physical Goldstone boson, called
majoron \cite{CMP}. In order to be consistent with the measurements of
the invisible $Z$ decay width at LEP the majoron should be (mostly) a
singlet under the \21 \gau symmetry.

Although the original majoron proposal was made in the framework of
the minimal seesaw model, and required the introduction of a
relatively high energy scale associated to the mass of the
right-handed \neus \cite{CMP}, there are many attractive theoretical
alternatives where lepton number is violated spontaneously at the weak
scale or lower. In this case although the majoron has very tiny
couplings to matter and the \gau bosons, it can have significant
couplings to the Higgs bosons.  As a result the Higgs boson may decay
with a substantial branching ratio into the invisible mode
\cite{JoshipuraValle92}
\begin{equation}
H \to J\;+\;J
\label{JJ}
\end{equation}
where $J$ denotes the majoron. The presence of this invisible decay
channel can affect the corresponding Higgs mass bounds in an important
way.

\begin{figure}[t]
\centerline{\protect\hbox{\psfig{file=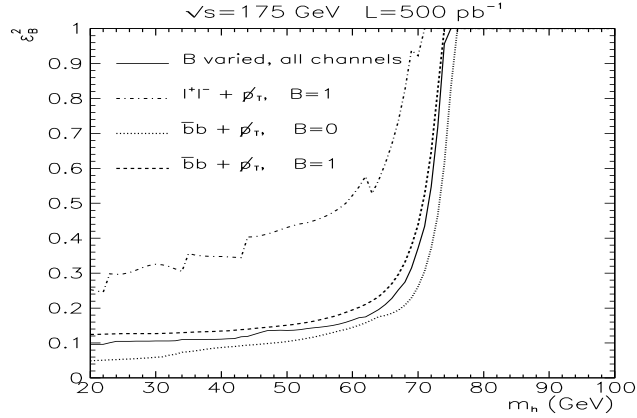,height=6cm,width=9cm}}}
\vglue -0.3cm
\caption{Region in the $\epsilon^2$ vs. $m_H$ that can be
excluded by LEP II.}
\label{ebolep}
\end{figure}

The production and subsequent decay of a Higgs boson which may decay
visibly or invisibly involves three independent parameters: its mass
$M_H$, its coupling strength to the Z, normalized by that of the
SM, $\epsilon^2$, and its invisible decay branching ratio.
The LEP searches for various exotic channels can be used in order to
determine the regions in parameter space that are already ruled out.
The exclusion contour in the plane $\epsilon^2$ vs. $M_H$, was shown
in \fig{ebolep} taken from ref. \cite{ebolep}.

\begin{figure}[t]
\centerline{\protect\hbox{\psfig{file=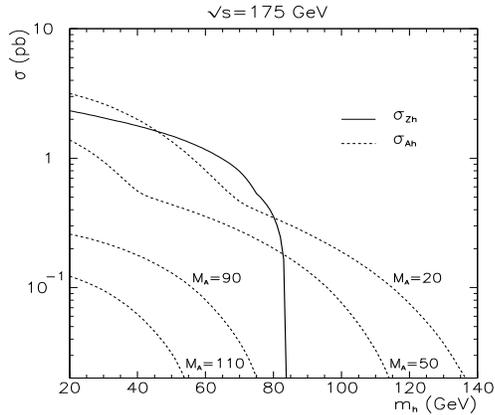,height=6cm,width=7cm}}}
\vglue -0.3cm
\caption{Total cross section for ZH and ZA production at LEP II.}
\label{ssig}
\end{figure}

\begin{figure}[t]
\centerline{\protect\hbox{\psfig{file=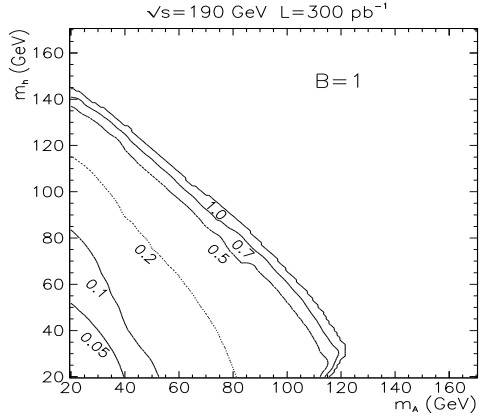,height=6cm,width=7cm}}}
\vglue -0.3cm
\caption{LEP II sensitivity on $\epsilon^2_{A}$ in the $m_A,m_H$
plane}
\label{iah_190}
\end{figure}

Another mode of production of invisibly decaying Higgs bosons is that
in which a CP even Higgs boson is produced at LEP in association with
a massive CP odd scalar \cite{ebolep}. This production mode is present
in all but the simplest majoron model containing just one complex
scalar singlet in addition to the SM Higgs doublet. As
seen in \fig{ssig}, the cross section for this is typically higher
than for the ZH channel, as long as $m_A$ is not too large.
Present limits on the relevant parameters are given in \fig{iah_190}, 
taken from ref. \cite{ebolep}.  In this plot we have assumed BR
($H \to J\:J$) = 100\% and a visibly decaying A boson.
Similar analysis were made for the case of a high energy linear $e^+
e^-$ collider (NLC) \cite{EE500}, as well as for the LHC \cite{granada}.

Before concluding let me mention that there are many novel properties
of SUSY Higgs bosons, squarks and sleptons that have recently been
explored and which may have important implications for the
corresponding searches at colliders. For a set of recent references
see \cite{Hsf}.

\subsection{Outlook}
\vskip .1cm

\ben
\item
There are good hints from experiment that neutrinos may be massive, a
possibility which is very attractive from the theoretical point of
view. This opens the way to many signatures such as neutrino
oscillations, neutrino-less double beta decays, possible distortions
in beta decay spectra and lepton flavour violating processes.
\item
The theoretical attractiveness of supersymmetry justifies the effort
devoted to the associated physics and its possible manifestations at
present and future particle colliders. So far the negative searches
for supersymmetric particles explore little of the relevant region and
rely strongly on the {\sl ad hoc} assumption of R--parity
conservation.
\item
There is a wealth of phenomena that could be associated both to the
physics of neutrino mass and to SUSY. They cover a very broad range of
energies and experimental situations. In this talk I have considered
two examples:
\bi
\item
models where neutrino masses are generated through lepton number
violation at the weak scale
\item
R--parity violation as the origin of neutrino mass. 
\ei
These two classes of schemes are theoretically attractive and lead to
a plethora of processes that could be seen in the new generation of
particle colliders, from the present LEP II and Tevatron to the future
LHC and NLC. R--parity violation provides a hybrid model for \neu mass
generation, combining radiative and seesaw ideas. One finds that \mnt
is naturally suppressed and calculable in terms of SUSY parameters and
the b-quark Yukawa coupling, while the magnitude of R--parity
violation at colliders can be large.
\een
In short, detecting neutrino masses is one of the main challenges in
particle physics, with far-reaching implications also for the
understanding of fundamental issues in astrophysics and cosmology. On
the other hand, probing Higgs boson physics and searching for SUSY in
whatever form is undoubtedly the main goal in the agenda of the next
generation of experiments, from elementary particle colliders down to
underground experiments and \neu telescopes.  We should be prepared for
exciting times where physics beyond the desert will show up!

\section*{Acknowledgements}

Supported by DGICYT grant PB95-1077 and in part by EEC under the TMR
contract ERBFMRX-CT96-0090.

\end{document}

\bibliographystyle{ansrt}

\end{document}